# Study of optical properties of x-ray system based on two zone plates.

A. Kuyumchyan [1*], A. Isoyan [1*], V. Kohn [2], I. Snigireva [3], A. Snigirev [3], S. Kuznetsov [1], V. Aristov[1], E. Shulakov [1]

[1] IMT RAS, 142432, Chernogolovka, Moskow District, Russia; [2] RRC "Kurchatov Institute", 123182, Moscow, Russia; [3] ESRF, BP-220, F-38043, Grenoble, France

We presents the results of study of focusing and imaging properties of double-lens system for hard x-ray radiation consisting of two Fresnel zone plates (ZP) made from silicon.

We demonstrate for the first time the phenomenon of focusing by two crystal ZP located at significant distance from each other. We investigate by both theoretically and experimentally the peculiarities of intensity distribution at the focal plane during a scan by second ZP normally to the optical axis of the system. We investigate as well the intensity distribution along the optical axis for our double-lens system from crystal ZP.

We realize experimentally a registration of the focused image of the object by means of double-lens system based on ZP. Measurements are performed on the beam line BM-5 of the European Synchrotron Radiation Facility (ESRF) at the x-ray energy 9.4 keV. We elaborate a computer program for theoretical simulation of the optical properties of x-ray double-lens system based on ZPs. A calculation is made by convolution of transmission function and Kirchhoff propagator in paraxial approximation by means of Fast Fourier Transformation.

## Introduction

Further development of x-ray optics significantly depends on creating high efficiency and simultaneously high resolution lens. Up to now there is a problem to make such lens for hard x rays. A high resolution can be reached by means of Fresnel zone plate (ZP) but efficiency of such lens is small as a rule [1-4]. With a usage of synchrotron radiation even the efficiency of few percents is sufficient for successful experiment, particularly, for obtaining the focused image of the object. In the report [1] authors used ZP from tantalum with zone thickness of 200 nm and the last zone width of 50 nm. The resolution of such ZP is equal to 200 nm, but the efficiency is only 1.5%. Nevertheless it was sufficient for obtaining the focused image. In the report [2] the multiple step ZP was performed by means of electron lithography and the efficiency ~ 70% was achieved. However, the resolution of such ZP becomes significantly smaller than the theoretical limit due to diffraction on the steps. In the report [3] ZP from gold was performed by means of deep x-ray lithography with zone thickness of 3.5 μm and the last zone width of 50 nm. As was noted by authors, the structure of zones with the width smaller than 200 nm is deviated randomly from the correct form. Moreover, an absorption of x rays on the gold slab of 3.5 μm width is not small for the x-ray energy 8 keV (the transmission is only ~ 30%). Therefore to obtain the theoretical limit of efficiency 40.5% is impossible in principle. In the report [4] the efficiency of 10% for the energy of 0,39 keV was achieved with ZP from Ni having the aperture of 120 μm and the last zone width of 30 nm. In the report [5] to increase the efficiency two ZPs from gold were located on small distance between them. As a result the efficiency of 19% was achieved.

We note the low efficiency ZP can be used only with synchrotron radiation. For laboratory source we need to have the efficiency about few tens percents. In works [6,7] ZP from silicon mono-crystal for axis geometry having 39% efficiency was prepared for the first time by means of electron lithography and ion plasma etching. In experiment on the beam line BL29XU of the synchrotron source SPring-8 (Japan) the resolution of this lens was measured by means of knife scan, which good corresponds to the theoretical estimation $\delta = 1.22 \Delta r$. In work [8] crystal ZP was used to obtain the focus image of the phase micro-object. This object consists of letters of small height making on the even surface of silicon crystal.

For creating x-ray optics devices with new properties it is significantly important to develop multiple lens systems. In this work we present the results of both experimental and theoretical study of focusing and imaging properties of double-lens system for hard x ray radiation based on the silicon crystal ZP.

## 2. The experimental setup and results of measurements

The experimental study with two zone plates was made on the synchrotron source ESRF (Grenoble, France), on the beam line BM-5 (beam from bending magnet). Let us denote the first zone plate by ZP#1, and the second zone plate by ZP#2.

We consider two different experimental setups. The first setup was used for investigation of focusing properties of two ZP made from silicon at the X-ray energy 9.4 keV (Fig. 1a). The second setup was used for obtaining the focused image of the optical fiber at the same x-ray energy (Fig. 1b).

In both cases the synchrotron radiation beam having a vertical size of 80 μm was made monochromatic by means of two-crystal monochromator Si (111). The total distance from the source to the entrance slit was equal to 40 m. The size of the slit is equal to 350x350 μm$^2$. In the first case the beam after the slit falls normally on the first zone plate ZP#1. The distance from the slit to ZP#1 is equal to 30 cm. The second zone plate ZP#2 was located at the distance 39 cm after ZP#1. The zone plate can be adjusted relative the optical axis with the accuracy 100 nm. We used two similar ZPs made from silicon mono-crystal with the following parameters: the radius of the first zone 14.38 μm, the number of zones 324, the last zone width 0.4 μm, the aperture 518 μm, the focal length 157 cm, the relief height 6 μm, The thickness of the membrane 2 μm. The transmission of the membrane for the x-ray energy 9.4 keV is equal to 93%.

The mean transmission of the surface relief 91%. The raster electron microscopy (REM) image and schematic image of ZP are shown in Fig.2. The intensity distribution is registered by means of FReLoN camera, consisting of specialized CCD detector. The space resolution of the FReLoN camera is about 1 μm. During the measurement the distance to detector was changed with small step.

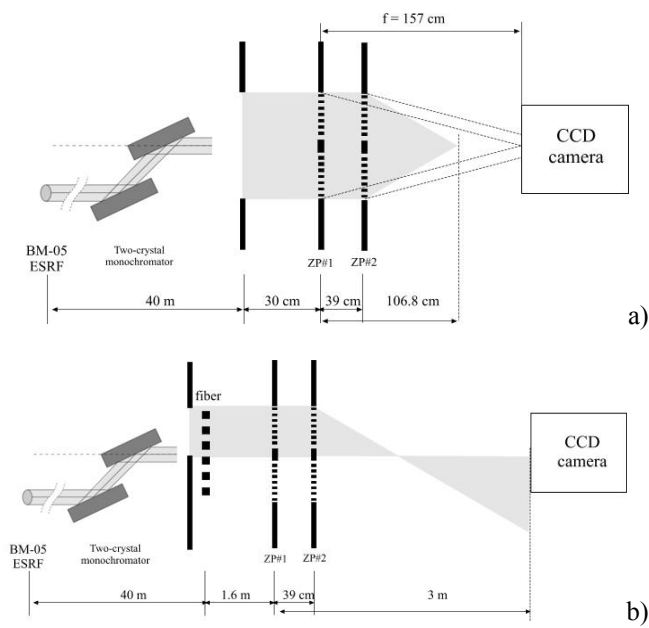

Fig.1. The experimental setup for: a) study of source focusing; b) obtaining the focused image of micro-object.

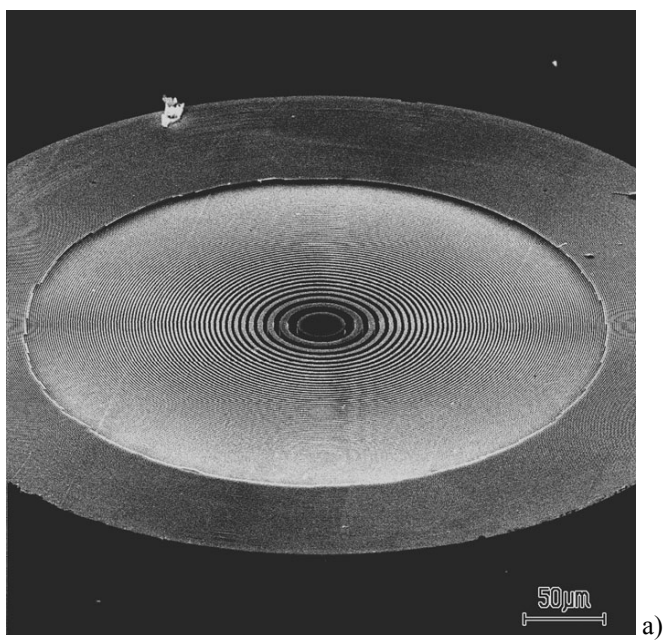

Fig.2. (a) REM image of Fresnel zone plate made from silicon, (b), schematic image of ZP.

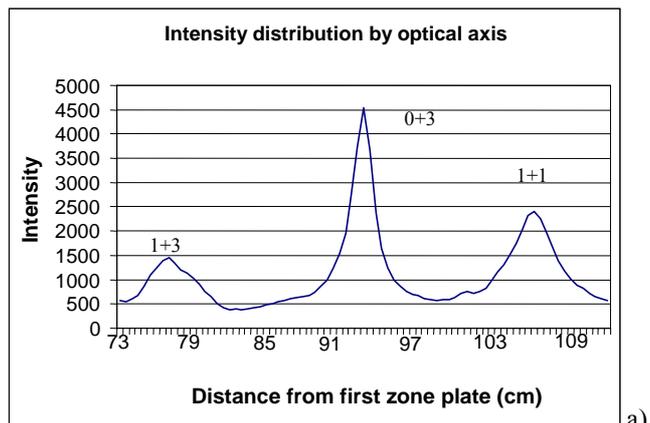

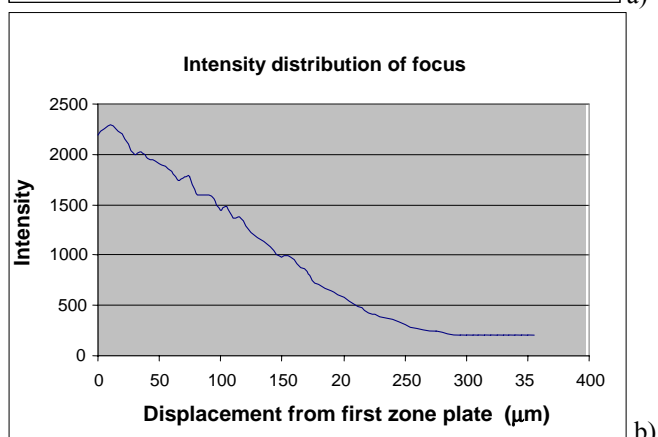

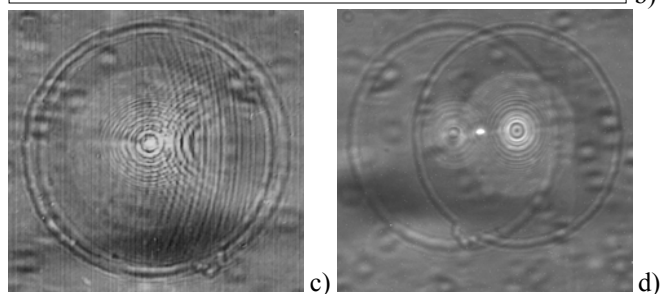

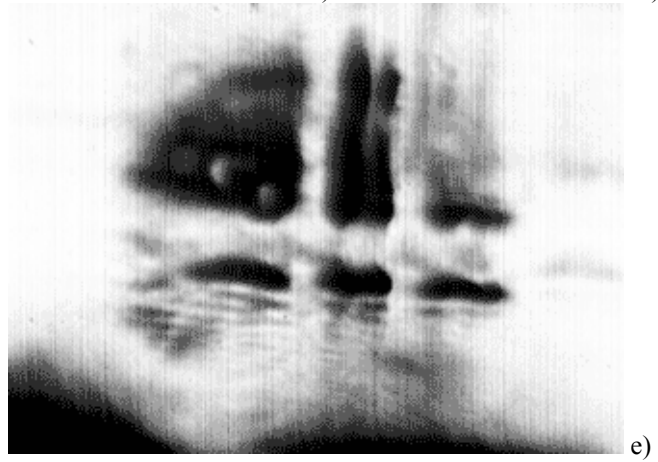

Fig.3. Measured intensity distribution along and across to optical axis in various cases (see text for details).

Fig. 3a shows a distribution of the integral intensity along the optical axis after the second lens ZP#2 within the region of distances from 73 cm to 115 cm counted between the first lens and the detector. One can see, the system has three most effective focuses which are formed by the way described below.

The first focus is registered at the distance 77 cm and it is formed by the first order of ZP#1 and third order of ZP#2

(we note it as 1+3). The second focus is registered at the distance 93 cm and it is formed by zero order of ZP#1 and third order of ZP#2 (0+3). The third focus is registered at the distance 107.5 cm and it is formed by the first order of ZP#1 and first order ZP#2 (1+1).

Fig. 3b shows how the focal intensity is changed in the focal plane 1+1 when the second lens moves inside the plane (normally to the optical axis). In this case the FReLoN camera is kept at fixed position and only the location of the second lens relative the optical axis is changed.

We show two images (Fig. 3c и 3d) which corresponds the displacements 10 μm and 100 μm. One can see in Fig. 3c that for small displacement of the second ZP the image shows a moiré pattern that allows one to adjust ZP#1 and ZP#2 relative the optical axis with accuracy up to 100 nm. The dependence of the focus intensity on the displacement of the second lens normally to optical axis becomes not monotone as shown in Fig.3b. According to the experimental results the maximum of focus intensity (1+1) arises at the displacement about few microns.

We measured the focus sizes for all three distances of focusing by the system of two ZPs in our experiment using the knife scan method. For (1+1) the focus size is equal to 2 μm; for (0+3) it is equal to 0.7 μm; for (1+3) it is equal to 1.1 μm. We note these sizes depend significantly on the source size. The source size influences as well the efficiency.

In the second experimental setup the beam after the slit falls on the test sample consisting of two optical fibers having diameter of 40 μm and keeping in cross geometry. The obtained image turns out to be increased twice as well as twice repeated (see Fig. 3e).

## 3. The theory and computer simulations

As is known [9], there exists analytical theory for the focal spot structure at the focal plane for one zone plate. Both cases of circular zones (for 2D-focusing) and linear zones (for 1D-focusing) allows analytical description. As for the system of two circular ZPs located at significant distance between them, the analytical calculations are absent and, probably, they are impossible in principle. Nevertheless, it is possible to assume that, at least, focus distances for such system can be the same as for two thin parabolic refractive lenses. To obtain the formula similar to the lens formula for the system of two ZPs, we calculate the ray trajectory from the point source through two lenses up to the point of intersection with the optical axis.

Let us denote the distance from the source to the first lens as $r_o$, the distance from the first lens to the second lens as $L$ and the distance from the second lens to the image plane as $r_i$. Let the focal length for the first lens be $F_1$, and for the second lens– $F_2$. As a result we obtain the formula which can be written in the following symmetric form

$$A_1 + A_2 - LA_1A_2 = 0, \quad A_1 = \frac{1}{F_1} - \frac{1}{r_o}, \quad A_2 = \frac{1}{F_2} - \frac{1}{r_i}. \quad (1)$$

It is easy to verify that the formula satisfies to the reciprocity principle because after a replacement source by detector, first lens by second lens and vice versa the formula stays the same. It is useful for our purpose to write this formula as equality for the distance $r_i$, namely,

$$r_i = F_2 \frac{1 - LA_1}{1 + (F_2 - L)A_1}, \quad A_1 = \frac{1}{F_1} - \frac{1}{r_o}. \quad (2)$$

For the zone plates used in the experiment we obtain $F_1 = F_2 = \rho_1^2/\lambda n = 1.57/n$ m, where $\rho_1 = 14.38$ μm is the radius of the first Fresnel zone, $\lambda = 0.132$ nm is the wavelength of x rays and $n$ is the order of focusing. Applying the formula (2) for $r_o = 40.3$ m, $L = 0.39$ m and various orders of focusing we obtain $r_i$ as 0.368 m (1+3), 0.529 m (0+3) и 0.693 m (1+1) that is very close to the values 0.38 m, 0.54 m and 0.685 m, which were obtained in experiment. Thus, it is verified that for rather long distance between them the system of two zone plates shows focusing properties which are rather closed to these for refracting lenses with the same parameters. Nevertheless, the small difference takes place. On the other hand, it is evident, that for small distance between the zone plates and in the limit of zero distance they must work as one zone plate with the doubled relief height and be quite different from refractive lenses. Probably there is some lower limit on the distance between the zone plates where the formula (2) still works. This interesting question will be investigated in the subsequent works.

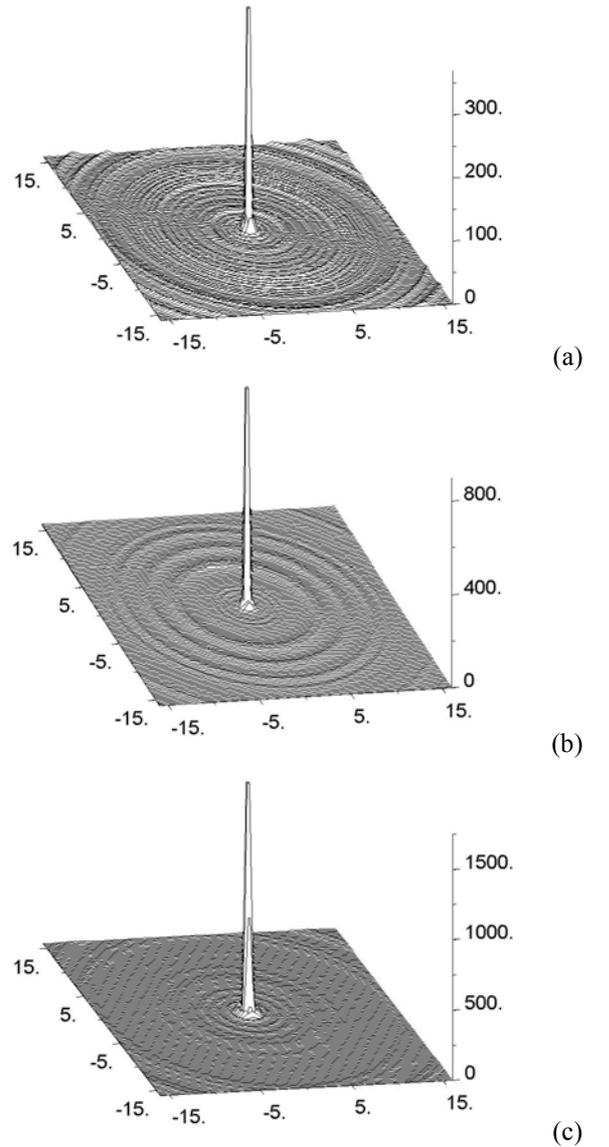

Fig.4. Theoretical intensity distribution for the wave focused by two zone plates and emerged from the point source: (a) focusing 1+3; (b) focusing 0+3; (c) focusing 1+1. Transverse distances are measured in microns, intensity is related to the value without zone plate focusing.

We have elaborated the computer program for simulating the intensity distribution at the plane perpendicular to the optical axis due to focusing a radiation from the point source by system of two zone plates. The program allows one to calculate a transmission of the transverse intensity distribution through the empty space by means of a convolution of the known complex wave field with the Kirchhoff propagator in the paraxial approximation. A calculation is performed by means of two Fourier transformations with a usage of the algorithm of Fast Fourier Transformation (FFT), because the Fourier image of the Kirchhoff propagator has a simple analytical form. Subsequently making a calculation for the first lens and then for second lens, we obtain the intensity distribution at any distance after the second lens.

Fig. 4 shows calculated intensity distributions for the wave in the planes normal to the optical axis at the distances, pointed above for the focusing of point source. Due to impossibility to have a sufficiently small grid of dots, a calculation was made for the zone plates having 40 zones. The used grid of dots has size 1024x1024 and step of 0.25 μm. The figure shows only the central part of the total region with the size 32x32 μm. The intensity is shown relative to its value at the same place in the case of absence of zone plates. As it follows from the calculation the largest intensity gain is reached for the case 1+1, that is rather understandable. However, in the experiment the integral intensity for this case is smaller than in the case 0+3 (see Fig. 3a).

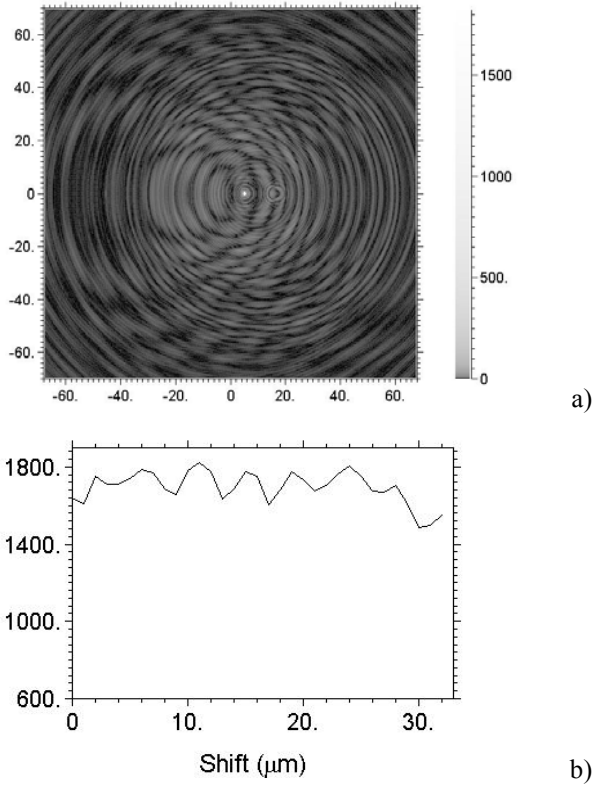

a)

b)

Fig.5. a) Moire pattern at the focus plane when the second lens is shifted on 11 μm; b) Dependence of peak intensity on the displacement. See text for details.

It is of interest to compare the maximum value of intensity for two zone plate with the same value for one zone plate. As follows from the theory for one zone plate with circular zones, the maximum intensity in the first order is estimated as $4N^2\sin^2(\varphi/2) = 3200$ for $N = 40$, $\varphi = 1.58$, if one neglects small absorption. On the other hand, for two lenses in the focus (1+1) the maximum intensity is equal to 1640, i.e. twice smaller. For focuses (0+3) and (1+3) we have correspondingly 940 and 350. However, an advantage of the two-lens system is a possibility to reduce focus distance from 1.6 m to 1.1 m counting from the first lens.

A calculation of intensity distribution at small transverse shift of the second lens relative the first lens leads, in general, to the same results as the experiment. As follows from the geometrical optics for the refractive lens, the transverse shift of the second lens on the distance $s$ leads to a shift of the focus point on the distance $d = sr_i/F_2$. In our case we have for the case (1+1) the formula $s = 0.44d$. On the same time the width and height of the intensity peak stay the same. For two zone plates a picture becomes more complicative. At small shifts of the second zone plate relative to the optical axis the focus point is shifted according to the formula, however the height of the intensity peak oscillates with 10 percent amplitude and in the points of maximums exceeds on 10 percent the value for well aligned second ZP.

Fig. 5b shows dependence of peak intensity on displacement of the second zone plate calculated with the step of 1 μm. In this case the position of the focus point is displaced as well, and characteristic moire pattern appears inside the region of aperture near the focus where the mean intensity is lower than a background intensity. It is of interest that the peak intensity not only does not decrease, but a mean value even slightly increase. Simultaneously it oscillates with rather small step. This phenomenon demands more detailed investigation which can be done in future work..

Fig. 5(a) shows a characteristic example of moiré pattern appeared for second lens displacement of 11 μm, i.e. at the point of maximum peak intensity. The intensity is shown in logarithmic scale, and the gray level g depends on intensity f according to the law $g = 1 - \log(f/f_{min})/\log(f_{max}/f_{min})$ under the condition that the values are restricted by the interval (0.1). The value $f_{min}$ may be slightly changed to be larger than the real minimum intensity. It is important if the latter is equal to zero. On the other hand, one can decrease the maximum value. The right scale shows a real correspondence between the gray level and the intensity value. As it follows from the figure, a displacement of second zone plate leads to appearance of extra focus of lower intensity at the right, and this focus becomes a center of new set of circular lines. At larger displacement new extra focus will appear at the left. Although their intensity are smaller compared to the main focus, their appearance is a new phenomenon which is unknown in the theory. This phenomenon is well observable in the experiment (see Fig. 3d), and it can be used for both adjusting the zone plates and study their quality.

## 4. Conclusion

It was shown both experimentally and theoretically that the system of two zone plates located at significantly large distance between them is able to focus the synchrotron radiation beam similarly the set of two refractive lenses. Both the formula for two lens on the focus distance and the formula for the focus displacement turn out to be valid. On the other hand, at small distance, evidently, the set of two zone plates is not similar to the set of two refractive lenses. Our study allows us to make a conclusion, that a usage of two zone plates allows one to decrease the focus distance. This may be useful for many applications. The obtained results show as well, that the study of the set of two zone plates must be continued because this system show new peculiarities which are absent for two refractive lenses.

\* \* \*

This work is supported by RFBR grants No. 03-02-17267, 04-02-17365. V. Kohn thanks a support of RFBR grants No. 03-02-16971, 04-02-17363, 05-02-16702.

Place figures in the text where you wish them to appear. They should be numbered consecutively in Arabic numerals and have suitable, self-explanatory captions. The width of lines and the size of the lettering should be large enough for the height of letters to be at least 1.5-2.0 mm.

\*e-mail: Arkuyumchyan@mtu-net.ru; aisoyan@mail.ru